\documentclass[prl,amsmath,amssymb,lengthcheck,showpacs,floatfix,groupedaddress,superscriptaddress,longbibliography]{revtex4-1}
\usepackage{graphicx}
\usepackage[english]{babel}
\usepackage{times}
\usepackage{dcolumn}
\usepackage[usenames,dvipsnames]{color}
\usepackage{units}
\usepackage[utf8]{inputenc}
\usepackage{bm}
\usepackage[normalem]{ulem}

\begin{document}

\title{Mottness collapse without metallisation in the domain walls of
triangular-lattice Mott insulator 1T-TaS$_2$}

\author{Jan Skolimowski}

\affiliation{Jo\v{z}ef Stefan Institute, Jamova 39, SI-1000 Ljubljana, Slovenia}

\author{Yaroslav Gerasimenko}

\affiliation{CENN Nanocenter, Jamova 39, SI-1000 Ljubljana, Slovenia}

\author{Rok \v{Z}itko}

\affiliation{Jo\v{z}ef Stefan Institute, Jamova 39, SI-1000 Ljubljana, Slovenia}
\affiliation{Faculty  of Mathematics and Physics, University of Ljubljana,
Jadranska 19, SI-1000 Ljubljana, Slovenia}

\date{\today}

\begin{abstract}
1T-TaS$_2$ is a charge-density-wave (CDW) compound with a
Mott-insulating ground state. The metallic state obtained by doping,
substitution or pulsed charge injection is characterized by an
emergent CDW domain wall network, while single domain walls can be
found in the pristine Mott state. Here we study whether and how the
single walls become metallic. Tunneling spectroscopy reveals
partial suppression of the Mott gap and the presence of in-gap states
strongly localized at the domain-wall sites. Using the real-space
dynamical mean field theory description of the strongly correlated
quantum-paramagnet ground state we show that the local gap suppression
follows from the increased hopping along the connected zig-zag chain
of lattice sites forming the domain wall, and that full metallisation
is preempted by the splitting of the quasiparticle band into bonding
and antibonding sub-bands due to the structural dimerization of the
wall, explaining the presence of the in-gap states and the low density of
states at the Fermi level.
\end{abstract}

\maketitle

\newcommand{\vc}[1]{{\mathbf{#1}}}
\renewcommand{\Im}{\mathrm{Im}}
\renewcommand{\Re}{\mathrm{Re}}

\newcommand{\expv}[1]{\langle #1 \rangle}
\newcommand{\ket}[1]{| #1 \rangle}
\newcommand{\Tr}{\mathrm{Tr}}

\newcommand{\yg}[1]{{\color{green} #1 }}

The interplay between superconductivity (SC) and correlated insulating
phases, such as Mott insulators and charge density waves (CDW), is one
of the central problems in condensed-matter physics. Remarkably, their
combination can be found even in simple material systems, such as
transition metal dichalcogenide (TMD) van der Waals compound
1T-TaS$_2$. The ground state is a Mott
insulator\cite{tosatti1976,fazekas1979,fazekas1980} with
CDW\cite{thomson1988} and unconventional quantum spin liquid behavior
\cite{anderson1973,klanjsek2017,law2017,kratochvilova2017,ribak2017,yu2017}.
Upon Se substitution\cite{liu2013}, Fe intercalation\cite{li2012}, or
by applying pressure\cite{sipos2008} it becomes superconducting, with
both CDW and correlated behavior still present. Further control over
electronic properties is possible through non-equilibrium charge
injection via ultrafast optical or electrical pulses
\cite{stojchevska2014,yoshida2015,vaskivskyi2016,ma2016,cho2016,svetin2017,gerasimenko2017,gerasimenko2018},
which lead to drastic insulator to metastable metal transition. The
long-standing hypothesis for metallisation and SC onset is linked to
the formation of CDW domain walls, seen in multiple TMDs with
different techniques\cite{joe2014, yan2017, qiao2017}. Recently, it
was challenged experimentally with scanning tunneling spectroscopy
(STS) \cite{cho2017}, which showed the absence of metallisation in
certain types of walls. First-principles calculations revealed that
atomic reconstruction in the walls may cause the formation of bound
states\cite{cho2017} and band reconstruction\cite{qiao2017}, but the
correlation effects were left out of scope. Thus, the crucial question
of whether the CDW distortion inside the wall can lead to Mottness
collapse remains open. In this paper we combine STS and dynamical mean
field theory (DMFT) calculations to study the behavior of the Mott gap
in CDW domain walls, finding Mottness collapse without metallisation.

\begin{figure}[htbp]
\begin{center}
\includegraphics[width=0.99\columnwidth]{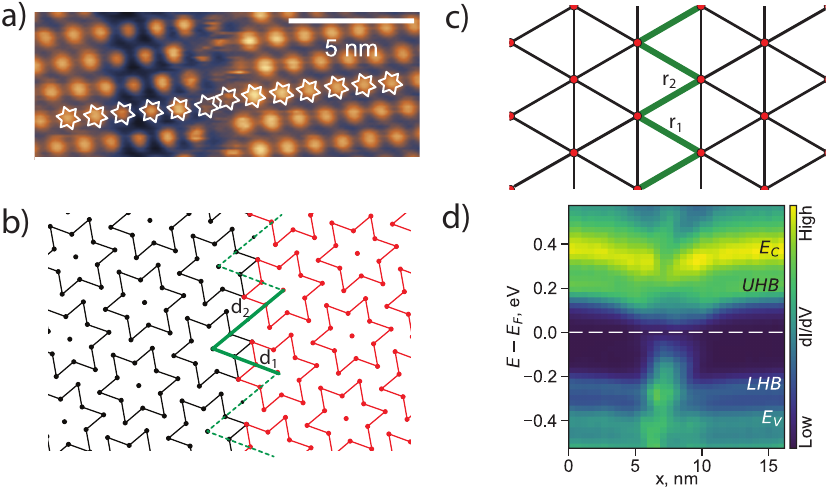}
\caption{Domain walls in 1T-TaS$_2$. (a) STM topographic image of a
single domain wall separating two adjacent CDW domains
($V_\mathrm{tip}=-\unit[0.8]{V}$, $I=\unit[100]{pA}$,
$T=\unit[4.2]{K}$). (b) Schematic diagram of the configuration of
David stars near the domain wall: the stars along the wall are brought
closer together. The modified center-to-center distances are indicated
as $d_1$, $d_2$. (c) Hubbard model description of the domain wall:
one-dimensional zig-zag chain with modified hoppings (ratios $r_1$,
$r_2$, green line) embedded in a two-dimensional Mott insulator
environment. (d) $dI/dV$ map across the domain wall (domain wall A,
DW-A, averaged over longitudinal direction). LHB and UHB are lower and
upper Hubbard bands, $E_c$ and $E_v$ indicate the additional
conducting and valence bands associated with the Ta $d$-shell.}
\label{Fig0}
\end{center}
\end{figure}

\begin{figure}[htbp]
\begin{center}
\includegraphics[width=0.99\columnwidth]{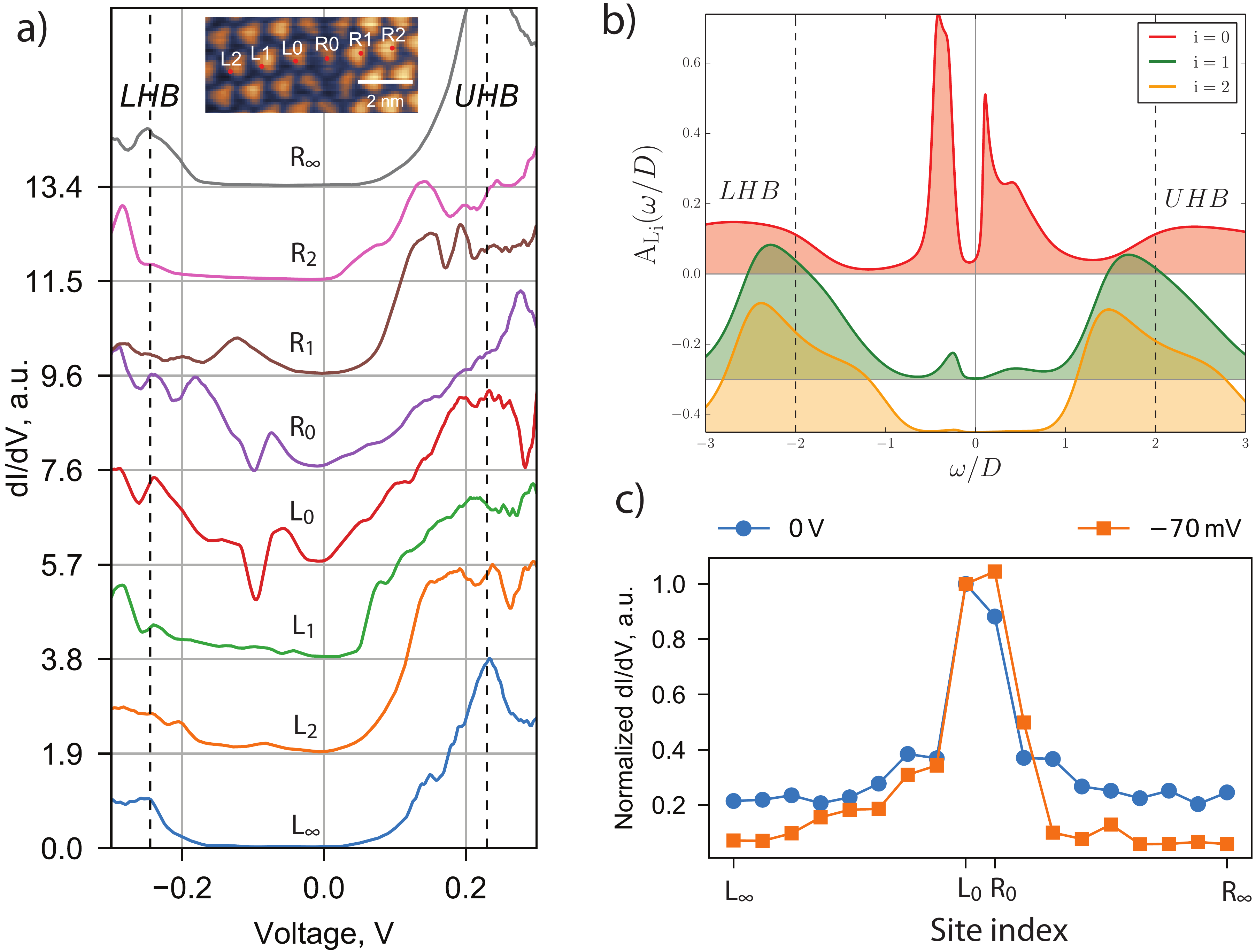}
\caption{Sub-gap features in the domain wall. (a) Position-resolved
$dI/dV$ tunneling spectra across the domain wall B (DW-B). L$_i$ and
R$_i$ indicate left and right side; $i$ counts the distance from the
domain wall centre in units of supercells. The spectra are offset for
clarity, the horizontal grid lines correspond to $dI/dV=0$ level for
each curve.  (b) Theoretical prediction for the spectra at the domain
wall and in its vicinity. (c) Cross-sections of spectra at selected
voltages: at Fermi level (blue line, $V=0$) and at the in-gap level
(orange line, $V=-\unit[70]{meV}$). Spectra are normalized to the
value at the point $L_1$.}
\label{figA}
\end{center}
\end{figure}

In 1T-TaS$_2$, each layer is periodically modulated to form a
$\sqrt{13}\times\sqrt{13}$ superlattice of David star deformations
\cite{thomson1988}, resulting in a commensurate CDW state with a
single half-filled electron band at the Fermi level
\cite{rossnagel2011,darancet2014}. The Coulomb repulsion opens a
charge gap in this band, resulting in a Mott insulating ground state
\cite{fazekas1979,kim1994,perfetti2006,perfetti2008,ligges2017}.
Single CDW domain walls (DWs) \cite{ma2016,cho2017,karpov2018} can be
found connecting large lattice inhomogeneities in freshly cleaved
samples or they remain after the relaxation of the metastable metallic
mosaic state created with voltage pulse from the scanning tunneling
microscope (STM) tip. We study both examples and find the relevant
physics similar.

Topographic STM image, Fig.~\ref{Fig0}(a), reveals individual David
stars as bright spots packed in a triangular lattice. The DW is seen
as a misfit of the lattices on left and right sides: David stars
partially overlap and are shifted with respect to each other. Among
the twelve possible DW types only a few are observed
experimentally\cite{ma2016}: indeed, in the most commonly seen DWs,
the stars are drawn closer together, Fig.~\ref{Fig0}(b), but still
retain their shape; see Supplemental Material (SM) for an extended
discussion.  The DW is very localized, approximately two superlattice
constants in extent, so that its core roughly consists of a 1D zig-zag
chain of the nearest-neighbor stars, Fig.~\ref{Fig0}(c).

The tunneling spectra, Fig.~\ref{Fig0}(d), reveal the apparent Mott
gap closing at the DW, while the clear bending of the Hubbard bands
outside the DW indicates its charging. The CDW gap is quite robust,
except at the very center of the DW. Changes to the Mott gap occur on
significantly larger spatial scales, allowing us to decouple the two
gaps. The spectral weight is distributed unevenly along the DW (see
SM), suggesting finite scattering. Closer inspection of the
high-resolution density of states (DOS) in the Mott gap, done on
another DW sector, shows two more major features: (i) inside the DW,
the DOS is significantly reconstructed and a small non-zero weight
appears at the Fermi level, and (ii) in-gap states at \unit[-70]{mV}
are present on top of a smooth spectrum, see Fig.~\ref{figA}(a). 

We now construct an effective model to understand the observed
low-energy behavior in the DW. The minimal model for the Mott state in
TaS$_2$ is the single-band Hubbard model on a triangular lattice at
half filling
\cite{imada1998,gebhardt,aryanpour2006,perfetti2005,perfetti2006,perfetti2008,freericks2009tas2}:
\begin{equation}
\label{Ham}
H=\sum_{\langle i,j\rangle,\sigma} t_{ij} c^\dag_{i\sigma} c_{j\sigma}
+ \sum_{i\sigma} \epsilon_i n_{i\sigma} + \sum_i U_i n_{i\uparrow} n_{i\downarrow},
\end{equation}
where $i$,$j$ are site indices, $\sigma=\uparrow,\downarrow$ is spin,
$c^\dag_{i\sigma}$ are electron creation operators, $n_{i\sigma}=
c^\dag_{i\sigma} c_{i\sigma}$ are the local electron-occupancy
operators, $t_{ij}$ are the hopping constants between
nearest-neighbors, $\epsilon_i$ are the on-site potentials, and $U_i$
are the Hubbard repulsion parameters. Down to the lowest temperatures,
TaS$_2$ does not order magnetically \cite{klanjsek2017,ribak2017}. A
simple approximation to describe this experimentally observed
paramagnetic Mott insulating state, theoretically predicted to exist
in a range of $U$ between the paramagnetic metal and the $120^{\circ}$
Néel ordered state \cite{yoshioka2009,shirakawa2017}, is the DMFT
\cite{georges1996,perfetti2006}. This approach can be generalized to
the inhomogeneous case
\cite{potthoff1999prb1,potthoff1999prb2,potthoff1999,freericks2004,freericks,okamoto2004,helmes2008penetrate,jiang2012,lee2017,bakalov2016}.
Here we assume the DW lattice structure to be predetermined and fixed,
and we focus on the effects of such a deformation on the electronic
degrees of freedom.

The modified separation between the stars in the DW is reflected in a
change of the corresponding hopping constants $t_{ij}$ due to a
different overlap of the corresponding Wannier wavefunctions; see the
adjacent black and red stars in Fig.~\ref{Fig0}(b), represented by the
green line in Fig.~\ref{Fig0}(c). The rescale factor for the hopping
constants will be denoted as $r$. (In unperturbed bulk $t_{ij} \equiv
t$, while $t_{ij} =r t$ for $i,j$ belonging to the DW.) The
superlattice deformation along the DW has a longitudinal component
leading to a period-2 modulation of $r$ along the DW (the values will
be denoted by $r_1$ and $r_2$). Next, we allow for a modification of
the on-site energies $\epsilon_i$ resulting from the deformation of
the stars. The additional on-site potential at the DW sites will be
denoted as $\delta$. For simplicity we assume that the Hubbard
constants remain uniform, $U_i \equiv U$. We set $U=4D$, where $2D=9t$
is the bandwidth of the triangular-lattice DOS. We fix the chemical
potential in the centre of the bulk Mott gap. We now study the effects
of the inhomogeneity on the local DOS (LDOS) using the real-space DMFT
approach as described in SM. The main result is that in the DW the
Mott state collapses if at least one of $r_1, r_2$ is large enough,
due to an increase in the kinetic energy of the DW subsystem, however
the resulting state has a low Fermi-level LDOS for $r_1 \neq r_2$ due
to the splitting of the quasiparticle band into bonding and
antibonding subbands which tends to open a (pseudo)-gap.

\begin{figure}[htbp]
\begin{center}
\includegraphics[width=0.95\columnwidth]{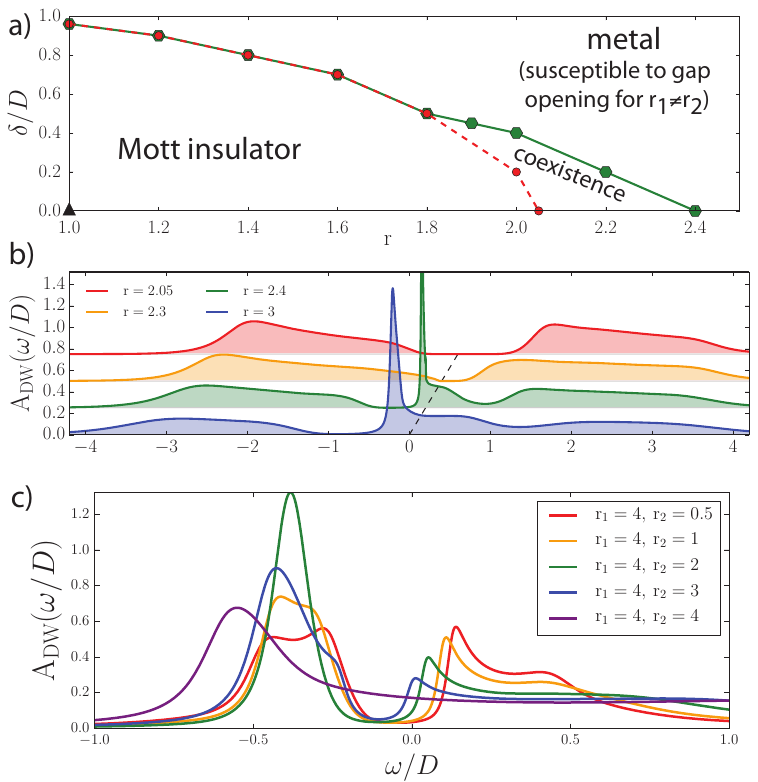}
\caption{Insulating and metallic regimes. (a) Phase diagram of the
domain-wall subsystem as a function of the hopping enhancement factor
$r \equiv r_1=r_2$ and the local potential $\delta$. (b) Domain-wall spectral
function across the bandwidth-driven Mott transition. (c) Effects of
the dimerisation, $r_1 \neq r_2$, on the quasiparticle band in a
metallic wall.}
\label{Fig1}
\end{center}
\end{figure}

In 1T-TaS$_2$, the lattice geometry effects are particularly
important: The zig-zag nature of the lattice distortion makes the DW a
connected linear system embedded in a bulk Mott insulator sheet. The
homogeneous Hubbard model undergoes Mott metal-insulator (MIT)
transitions of two types
\cite{imada1998,georges1992mott,zhang1993,georges1996,werner2007,zitko2013,logan2016}:
interaction/bandwidth-driven (Hubbard repulsion $U$ overcomes the
kinetic energy proportional to hopping) or doping-driven (for
sufficiently large $U$ the electron occupancy approaches
half-filling). If the system lacks translation invariance, the MIT can
also take place in a subsystem, e.g. on a surface, at an interface, or
in a domain-wall 
\cite{potthoff1999prb1,potthoff1999prb2,potthoff1999,freericks2004,freericks}.
In TaS$_2$, the DW sub-system metallises if $r_1=r_2 \equiv r$ becomes
large (bandwidth-driven MIT) or if $\delta$ becomes large
(doping-driven MIT), see Fig.~\ref{Fig1}(a).

\begin{figure}[htbp]
\begin{center}
\includegraphics[width=0.95\columnwidth]{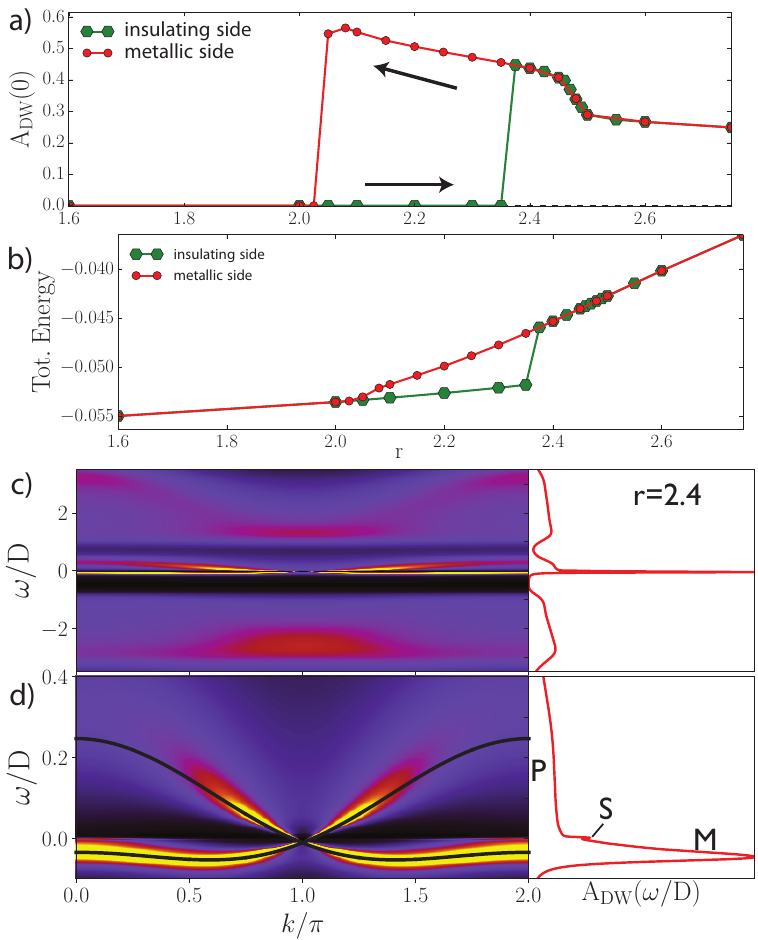}
\caption{Effects of the hopping enhancement factor $r$ with $\delta$
fixed at $0$.
(a) Local DOS at the Fermi level across the bandwidth-driven Mott
transition. The arrows in the hysteresis region indicate the direction
of the changing parameter $r$. (b) Total energies of metallic and
insulating solutions. 
(c,d)
Momentum-resolved spectral functions. $k$ is the momentum along the
DW. The black line is the dispersion of a non-interacting 1D zig-zag
chain tight-binding Hamiltonian. }
\label{Fig2}
\end{center}
\end{figure}

The ``local charge gap'' is reduced inside the DW. This is due to the
enhanced hopping $r$ that increases the effective bandwidth so that
local LHB and UHB become broader, while the band centers shift only little.
Because of the particle-hole asymmetry in the triangular lattice, the
top of LHB and the bottom of UHB have quite different shapes and the
upper edge of the LHB moves towards the Fermi level with increasing
$r$ faster than the bottom edge of the UHB. For high enough
$r=r_1=r_2$, the DW goes through a MIT and it metallizes: A
quasiparticle (QP) band emerges inside the Mott gap,
Fig.~\ref{Fig1}(b), so that the Fermi-level local LDOS jumps to a
finite value, Fig.~\ref{Fig2}(a). The QP band has a highly asymmetric
internal structure with a sharp maximum $M$ in the occupied part of
the spectrum (the maximum is not pinned to the Fermi level) and an
extended flat plateau $P$ in the unoccupied part, see Fig.~4(d), right
panel. In addition, in the range $r \lesssim 2.5$ we find a weak
secondary peak $S$ (see SM for its interpretation). Information about
the in-gap states can be extracted from the momentum-resolved
(partially Fourier transformed) spectral function, $A_{k,i}(\omega)$,
where $k$ is the longitudinal momentum in the DW direction, while $i$
is the index of the DW site in the transverse direction. An overview
is shown in Fig.~4(c), and a close-up to the QP band in Fig.~4(d). The
dominant features ($M$ and $P$) in the structure of the QP band are
due to the one-dimensional nature of the DW which is a 1D zig-zag
chain with nearest and next-nearest neighbour hopping constants
differing by a factor of $r$. The band dispersion for such a
non-interacting tight-binding model with $r=2.4$ indeed matches the QP
dispersion, Fig.~4(d).

Surprisingly, the boundary bandwidth-driven MIT is first order even at
zero temperature, unlike the bulk MIT. We find a sizeable region with
coexisting DMFT solutions: with increasing $r$, the insulating
solution metalizes at $r_{c,2}$, while with decreasing $r$ the
metallic solution becomes insulating at $r_{c,1}<r_{c,2}$. The
insulating solution is stable in the entire coexistance region,
$r_{c,1}<r<r_{c,2}$, Fig.~\ref{Fig2}(b). As a consequence, the
bandwidth-driven transition from the insulating state happens at the
moment when the Hubbard band (here LHB) touches the chemical
potential, similar to the scenario for the disappearance of the
metastable insulating DMFT solution in the bulk case.  We note,
however, that for $r>r_{c,2}$ the LHB becomes clearly separated from
the quasiparticle (QP) peak, i.e., at $r=r_{c,2}$ the LDOS changes
discontinuously. The doping-driven boundary Mott transition also
proceeds by the mechanism of LHB crossing the chemical potential, but
the QP peak is pinned to the top of the LHB right after the transition
(see SM for more details).

The superlattice deformation at the DW corresponds to an increase of
hopping that can easily attain values of $r \approx 3$ or $4$ (see SM
for estimates corresponding to common DW types). One would thus
generically expect to observe a strong metallisation and a prominent
peak at (or close to) the Fermi level. In reality, however, the DW is
naturally dimerized so that $r_1 \neq r_2$. For most DW types $r_1$
and $r_2$ can indeed be quite different. The Mott phase collapses at
the DW if the kinetic energy of the 1D subsystem is high enough. For
this to occur, it is sufficient for one of $r_1, r_2$ to be large,
even if the 1D subsystem in isolation from the triangular-lattice bulk
would actually be gapped (a band insulator). The alternation of
hoppings gives rise to molecular orbitals localized on the nearby
sites forming dimers, thereby splitting the QP band into bonding and
anti-bonding subbands. The resulting spectrum has a prominent peak in
the occupied part of the spectrum ($\omega<0)$ and a gap-like strong
suppression of DOS around the Fermi level, see Figs.~\ref{figA}(b) and
~\ref{Fig1}(c).

The major observations from the tunneling spectra can thus be
consistently understood within this model. In the DW core, two effects
dominate: reconstruction of spectra, and emergence of the bound state.
If the in-gap resonance is disregarded, it can still be observed that
the Mott gap is smaller and the DOS distribution is smeared - the
behavior reproduced by the intra-DW change of hopping. The in-gap DOS
peak at negative bias results from the QP band splitting. The features
in the unoccupied part of the the spectrum are most likely pushed
toward the UHB and are largely submerged into it. We also observe a
small yet clearly non-zero value of Fermi-level LDOS inside the DW,
which results from the subband spectral tails of the split QP band.
The spatial distribution of the Fermi level LDOS and the bound state
LDOS are indeed quite similar, see Fig.~\ref{figA}(c), which confirms
their common origin. Finally, the band bending outside the DW is
effectively reproduced with the local potential. Its role in
metallization is unpronounced inside the DW, even when UHB is brought
very close to the Fermi level (see SM). 

We thus conclude that the structural change in an isolated DW does not
yield a good metal in spite of the Mottness collapse. The same result
can be extrapolated to the DWs in the hidden state, which have an even
larger structural distortion \cite{ma2016, gerasimenko2017}. Optical
or electrical pulses can, however, affect the orbital structure
\cite{ritschel2015} or cause redistribution of electrons in the
Brillouin zone, and lead to metallization independently of the density
and the type of DWs. Finally, we mention the possibility that the
superconductivity in this compounds is unrelated to the DWs, and that
upon doping, the SC state emerges directly from the quantum spin
liquid state, thus 1T-TaS$_2$ might be an unconventional
resonanting-valence-bound superconductor.

\begin{acknowledgments}
We acknowledge the support of the Slovenian Research Agency (ARRS)
under P1-0044 and J1-7259, ERC AdG GA 320602 Trajectory, and
discussions with Peter Prelovšek, Jernej Mravlje, and Dragan
Mihailovic.
\end{acknowledgments}

\bibliography{domain-wall_10}

\end{document}